\documentclass[useAMS, usenatbib]{mn2e}
\usepackage{times}
\usepackage{graphicx}
\usepackage[usenames]{color}
\usepackage{ulem}
\newcommand{\mpc}{\mbox{$h^{-1}$Mpc}}
\newcommand{\mpci}{\mbox{$h$ Mpc$^{-1}$}}
\newcommand{\lyf}{Ly$\alpha$ forest}
\newcommand{\oned}{one-dimensional}
\newcommand{\thrd}{three-dimensional}

\begin{document}

\title[Flux Power Spectrum and Covariance]
{Ly$\bmath{\alpha}$ Flux Power Spectrum and Its Covariance}

\author[Zhan et al.]
{Hu Zhan,$^1$\thanks{E-mail: zhan@physics.ucdavis.edu} 
Romeel Dav\'{e}$^2$,
Daniel Eisenstein$^2$, and Neal Katz$^3$ \\
$^1$Department of Physics, University of California, 
    Davis, CA 95616, USA \\
$^2$Steward Observatory, University of Arizona, Tucson, AZ 85721, USA \\
$^3$Astronomy Department, University of Massachusetts at Amherst, 
MA 01003, USA }

\maketitle

\begin{abstract}

We analyze the flux power spectrum and its covariance using simulated
\lyf{}s. We find that pseudo-hydro techniques are good 
approximations of hydrodynamical simulations at high redshift. 
However, the pseudo-hydro techniques fail at low redshift because 
they are insufficient for characterizing some components of the 
low-redshift intergalactic medium, notably the warm-hot intergalactic 
medium. Hence, to use the low-redshift Ly$\alpha$ flux power spectrum
to constrain cosmology, one would need realistic hydrodynamical
simulations. By comparing (\oned{}) mass statistics 
with flux statistics, we show that the nonlinear transform between 
density and flux quenches the fluctuations so that the flux power 
spectrum is much less sensitive to cosmological parameters than the 
one-dimensional mass power spectrum. 
The covariance of the flux power spectrum is nearly Gaussian. 
As such, the uncertainties of the underlying mass power 
spectrum could still be large, even though the flux power spectrum can 
be precisely determined from a small number of lines of sight.

\end{abstract}

\begin{keywords}
cosmology: theory -- large-scale structure of universe
-- methods: numerical -- quasars: absorption lines
\end{keywords}

\section{Introduction}
The \lyf{} is a useful tool for studying the cosmic density 
field over a wide range of redshift that has not been
easily accessible to other types of observations. For each 
line of sight (LOS) to a quasar, one can sample the density field 
almost continuously in one dimension. With enough LOSs 
covering up to $z \sim 6$, the \lyf{} will enable us to 
establish a more complete picture of the universe and its evolution, and,
subsequently, to determine cosmological parameters. 

Statistics of the \lyf{} have been applied to many aspects of 
large-scale structure studies such as recovering the initial linear mass 
power spectrum \citep[PS;][]{cwk98, cwp99, h99, ff00, mmr00, msc04b, 
cwb02, gh02, zsh03}, measuring the flux PS and bispectrum 
\citep{hbs01, mms03, kvh04, vmh04}, estimating cosmological parameters 
\citep{mm99, zht01, cwb02, m03, smm03, vmt03, smm04, vwh04, lhh05},
inverting the \lyf{} \citep{nh99, pvr01, z03}, finding the 
applicable range of the hierarchical clustering model 
\citep{fpf01, zjf01}, and estimating the velocity field \citep{zf02}. 
These studies show that the \lyf{} has provided an important complement 
to studies based on galaxy and QSO samples.

In fact, the \lyf{} is becoming a major player in precision 
cosmology. It is one of the three measures (others being the cosmic 
microwave background and galaxy redshift surveys) in \citet{svp03} that
show a tantalizing hint of a possible running spectral index, which 
could have a significant impact on inflationary models. However, the 
degeneracy between the amplitude and slope of the flux PS that has arisen 
owing to uncertainties in the mean transmission weakens the argument for
a running spectral index \citep{smm03}. This motivates us to pursue a 
better understanding of the uncertainties in the Ly$\alpha$ flux 
statistics in the era of precision cosmology.

A number of authors have examined the effects of metal 
lines, continuum fitting, strong discrete absorptions systems, damping 
wings, ionizing radiation fluctuations, galactic winds, and simulation 
details \citep{hbs01, vhc04, mw04, msc04a, msc04b}. In this paper, we 
focus on the following facets of the \lyf{}. 
Firstly, we evaluate the performance of pseudo-hydro 
techniques \citep[e.g.][]{pmk95, cwk98} by comparing their results to 
those from full hydrodynamical simulations and by varying pseudo-hydro
parameters. We show that a systematic difference between pseudo-hydro
and full-hydro results exists, which might cause a bias in cosmological 
parameter estimation. Thus, careful calibrations of pseudo-hydro
techniques over a large dynamic range are necessary for precision 
cosmology.

Secondly, the Ly$\alpha$ flux is a nonlinear transform of the 
one-dimensional density field. Fluctuations in the flux and the sample 
variance of the flux PS are much smaller than those of the density field. 
Consequently, one can measure the flux PS to a high precision with a 
small number of LOSs, but to achieve the same precision in the mass PS
one needs a lot more LOSs.

Thirdly, the Ly$\alpha$ flux is nearly Gaussian, while the 
one-dimensional density field exhibits stark non-Gaussianity 
\citep{ze05}. These
are not characterized by the PS but by higher-order statistics such as 
the covariance of the PS. We quantify the correlation between Fourier 
modes in both flux and density fields using the covariance of the PS.

Finally, we point out the difficulty with extending the Ly$\alpha$
flux PS analysis to low redshift. Since the non-Gaussianity of the 
density field becomes stronger at lower redshift, one would need even 
more LOSs to recover the one-dimensional mass PS at the same precision 
that one would at high redshift. Meanwhile, pseudo-hydro techniques 
work poorly at low redshift so that one would have to utilize more 
time-consuming full-hydro simulations.

The rest of the paper is organized as follows. \S\ref{sec:sim} 
briefly describes the simulations used for the investigation. 
The flux PS and its covariance are illustrated in
\S\ref{sec:comp} and contrasted with \oned{} mass PS and its 
covariance. Comparisons of flux PSs in different cosmological models 
are given in \S\ref{sec:cos}, and \S\ref{sec:con} concludes
the paper. Note that in our terms the flux PS is always 
the \oned{} PS of the flux $F$, not $F/\bar{F}$.

\section{Simulated Ly$\bmath{\alpha}$ Forests} \label{sec:sim}

The \lyf{} probes deeply into the nonlinear regime of the cosmic
density field. This has made numerical simulations indispensable
for understanding the nature of the \lyf{} and inferring
cosmological parameters from flux statistics. Two types of 
cosmological simulations have been commonly used 
to simulate the \lyf{}. One is pure cold dark matter (CDM), or 
$N$-body simulations, which assume that baryons trace the dark matter 
\citep[e.g.][]{pmk95, cwk98, rpm98}. The other is hydrodynamical 
simulations \citep[e.g.][]{cmo94, zan95, hkw96, dhw97}. 
Although full hydrodynamical simulations are well suited for studies
of the \lyf{}, they are currently too time-consuming to explore 
a large cosmological parameter space as one often desires. Whereas, 
$N$-body simulations run much faster, and hence can be used to cover a 
wider range of cosmological models in a practical time. 
Other types of simulations such as hydro-particle-mesh simulations
\citep[HPM,][]{gh98} and the simple log-normal model \citep{bd97} have 
also been used to study the Ly$\alpha$ forest.

\subsection{Hydrodynamical Simulations with Photoionization}
\label{sec:hyd-lya}
We use a hydrodynamical simulation (HLCDM) for this study.
It is a variant of the
low-density-and-flat CDM (LCDM) model with a slight tilt of the 
initial power spectral index $n$ (see Table \ref{tab:para1}). 
HLCDM evolves $128^3$ CDM particles and $128^3$ gas particles from 
$z = 49$ to 0 using Parallel TreeSPH \citep{ddh97}. 
The box size is 22.222 \mpc{} in each 
dimension with a 5 \mbox{$h^{-1}$ kpc}
resolution. The simulation also includes star formation with feedback 
and photoionization \citep{kwh96}. The UV ionization background is 
from \citet{hm96}. 

We simplify the method in the software tool, {\sc tipsy}\footnote{
http://www-hpcc.astro.washington.edu/tools/tipsy/tipsy.html}, to 
produce \lyf{}s. The procedures are outlined below. For convenience, we
assume zero metallicity and do not include noise.
 
Snapshots of the simulations contain the position ${\bf r}_i$ and 
velocity ${\bf v}_i$ of each particle, where $i$ labels the
$i$th particle. Smooth-particle hydrodynamics (SPH) defines the baryon
density $\rho_{\rm b}({\bf x})$ at any location to be a sum of 
contributions from all nearby gas particles, i.e.
\begin{equation}
\rho_{\rm b}({\bf x}) = \sum_{i = 1}^{N_{\rm p}} 
m_i\, w(|{\bf x}-{\bf r}_i|, \epsilon^j_i),
\end{equation}
where $N_{\rm p}$ is the total number of particles, $w$ is the density 
kernel or the assignment function, $m_i$ is the mass of particle $i$, 
and $\epsilon^j_i$ is the smoothing length determined by the distance 
between particle $i$ and its $j$th neighbor ($j = 32$ in this paper). 
In practice, densities are assigned on a discrete grid for further 
analysis. We employ a spherically symmetric spline kernel from 
\citet{ml85}, which is also used in TreeSPH 
for force calculations. It has the form
\begin{equation}
w(r, \epsilon) = \frac{1}{\pi \epsilon^3}\left\{ \begin{array}{ll}
1 - \frac{3}{2} \left(\frac{r}{\epsilon}\right)^2 
+ \frac{3}{4} \left(\frac{r}{\epsilon}\right)^3 & 
\quad 0 \leq r < \epsilon \\ 
\frac{1}{4}\left[2- \frac{r}{\epsilon}\right]^3 & 
\quad \epsilon \leq r < 2 \epsilon \\ 
0 & \quad r \geq 2 \epsilon, \\ \end{array} \right.
\end{equation}
which vanishes beyond the radius $2\epsilon$ and has a smooth gradient 
everywhere. This density kernel is an effective 
low-pass filter that suppresses fluctuations on scales smaller than
$2\epsilon$ ($k > \pi/\epsilon$). 

For each gas particle we assume a universal hydrogen fraction 
of $0.76$ to convert the baryon density to hydrogen density, and 
calculate the ionization equilibrium H{\sc i} density at the particle 
temperature.
In principle, 
LOSs may be sampled in any random direction, but for computational 
simplicity we assign the H{\sc i} density on a grid of $256^3$ nodes,
and then extract one-dimensional fields randomly from this grid. 
We have tried a higher-resolution grid of $512^3$ nodes, and the 
results are not affected on scales above $0.6$ \mpc{}
($k < 10$ \mpci).
Node temperatures and velocities are also assigned as weighted 
averages of contributing particles. The weight is proportional to the
H{\sc i} mass contribution of each particle.

The assumption of ionization equilibrium certainly breaks down in 
very dynamic regions such as shocks. 
However, since the equilibrium H{\sc i} fraction calculated in such regions
is already considerably lower than elsewhere, there will not be much of an 
effect on simulated \lyf{}s, even if additional shock physics can 
further reduce the H{\sc i} fraction by orders of magnitude. In addition,
shock fronts, unlike shocked gas, only occupy a small fraction of the 
total simulation volume, so they could not have too much impact on 
the \lyf{}.

\begin{table}
\caption{Parameters of the simulations.}
\label{tab:para1}
\begin{tabular}{@{}lccccccc}
\hline
Model & Type & $\Omega$ & $\Omega_{\rm b}$ & $\Omega_\Lambda$ & $h$ & 
$n$ & $\sigma_\mathrm{8}$ \\
\hline 
HLCDM & Hydro. & 0.4 & 0.05 & 0.6 & 0.65 & 0.95 & 0.8 \\
TCDM  & $N$-Body & 0.3 & 0.04 & 0.7 & 0.7 & 1.1  & 0.8 \\
LCDM1 & $N$-Body & 0.3 & 0.04 & 0.7 & 0.7 & 1.0  & 0.8 \\
LCDM2 & $N$-Body & 0.3 & 0.04 & 0.7 & 0.7 & 1.0  & 1.0 \\
OCDM  & $N$-Body & 0.3 & 0.04 & 0   & 0.7 & 1.0  & 0.8 \\
\hline
\end{tabular}

\smallskip
With the exception of HLCDM, the baryon 
density is used only for generating the initial mass PS.
\end{table}

With the H{\sc i} density along the LOS, one 
can determine the Ly$\alpha$ optical depth $\tau$ and transmitted 
Ly$\alpha$ flux $F$ of each pixel (each node of the density grid). 
The mean flux $\bar{F}$ of the \lyf{} is constrained by 
observations. We adjust 
the intensity of the UV ionization background $\Gamma_{\rm UV}$ so that 
the mean flux of all pixels in the simulations follows
\begin{equation} \label{eq:mflx}
\bar{F}(z) \simeq \left\{ \begin{array}{ll}
\exp \left[-0.0032\,(1+z)^{3.37\pm 0.2}\right] & \ \ 1.5 \leq z \leq 4 \\
0.97-0.025\,z\pm (0.003+0.005\,z) & \ \ 0 \leq z < 1.5. \\
\end{array} \right.
\end{equation}
The high-redshift part of the mean flux formula is given by the 
observations of \citet{kcc02}, which is consistent with others 
\citep{lsw96, rms97, mmr00}. Since the mean opacity of the low-redshift 
\lyf{} is somewhat uncertain \citep{pss04}, we take the simulated mean 
flux from \citet{dhk99} as the fiducial mean flux at low redshift.
Thermal broadening is added afterward 
using the temperature of each pixel. Note that thermal 
broadening smoothes out small-scale fluctuations in the \lyf{} without 
altering the mean flux very much. Therefore, it preferentially reduces the 
flux power on small scales.

There is a slight
inconsistency in that HLCDM has already included the UV 
ionization background, yet we need to adjust the intensity of the UV 
radiation on outputs of the simulation to fit the mean flux. 
This inconsistency does not significantly affect the results that follow 
because the temperature of the intergalactic medium (IGM) is not 
sensitive to the UV background 
intensity \citep{cwk97}. In fact, the simulation outputs are able to 
reproduce the observed mean flux with their internal UV ionization 
background \citep{dhk99}. External adjustments are only needed to vary 
the mean flux within the given observational and numerical uncertainties.
Thus, even if the intensities of the externally adjusted UV 
background were used internally in the simulations, the LOS 
Ly$\alpha$ absorption would not change appreciably.

\begin{table}
\caption{Methods for generating the \lyf{}.}
\label{tab:psh}
\begin{tabular}{@{}lcccc}
\hline
Method & Particle & $\rho_{\rm b}$ & $T_{\rm node}$ & $\tau$ \\
\hline
HYDRO & SPH &  & SPH  & 	Ion. Eq. \\
BA-TE-IE & SPH &  & Thermal Eq.  & Ion. Eq. \\
DM-TE-IE & CDM & $\propto \rho_{\rm d}$ & Thermal Eq. & 
	Ion. Eq. \\
DM-EOS & CDM & $\propto \rho_{\rm d}$ & 
	$T_0 (\rho_{\rm b}/\bar{\rho}_{\rm b})^{\gamma-1}$ & 
	$\propto (\rho_{\rm b}/\bar{\rho}_{\rm b})^\beta$ \\
\hline
\end{tabular}

\smallskip
\end{table}

\subsection{Pseudo-Hydro Techniques}
\citet{pmk95} developed a pseudo-hydro technique for generating \lyf{}s 
from $N$-body simulations. They assume that the baryons trace the dark matter
and calculate the optical depths of baryons assuming ionization equilibrium. We further simplify their method by also assuming thermal 
equilibrium (labelled as DM-TE-IE in Table \ref{tab:psh}). 

Separately, \citet{cwk98} proposed a slightly different pseudo-hydro 
technique (labelled as DM-EOS). In addition to assuming baryons to trace 
dark matter, they also make use of the fact that, in thermal 
equilibrium, the equation of state (EOS) of the IGM gives rise to an 
approximate temperature--density relation
\begin{equation} \label{eq:eos}
T=T_0(\rho_{\rm b}/\bar{\rho}_{\rm b})^{\gamma-1}, 
\end{equation}
where $T_0 \sim 10^4$ K, $1.3\leq \gamma \leq 1.6$, and 
$\rho_{\rm b}/\bar{\rho}_{\rm b}\la 10$ \citep{hg97}. Since the 
Ly$\alpha$ optical depth is proportional to $\rho_{\rm b}^2\, T^{-0.7}$ in 
regions around the mean density, one finds
\begin{equation} \label{eq:lyaf}
F \simeq e^{-A (\rho_{\rm b}/\bar{\rho}_{\rm b})^\beta} 
\simeq e^{-A (\rho_{\rm d}/\bar{\rho}_{\rm d})^\beta},
\end{equation}
where $A \propto \Omega_{\rm b}^2\,\Gamma_{\rm UV}^{-1}T_0^{-0.7}$, 
$\beta = 2.7 - 0.7 \gamma$, and $\rho_{\rm d}$ is the dark matter density.
The constant $A$ is often left as a fitting parameter adjusted to 
reproduce the observed mean flux.

\begin{figure}
\centering
\includegraphics[width=84mm]{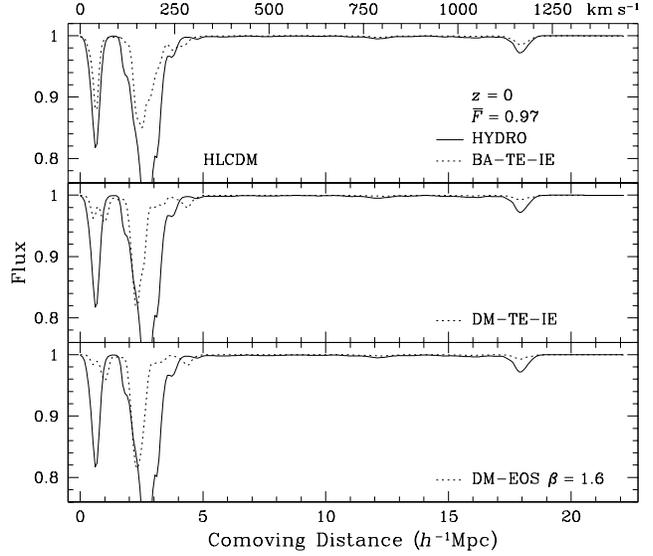}
\caption{Transmitted Ly$\alpha$ flux based on baryon and dark matter 
distributions at $z = 0$. The three panels compare \lyf{}s generated 
from the same LOS with methods HYDRO, BA-TE-IE, DM-TE-IE, and DM-EOS. 
All four methods are required to reproduce the same mean flux of 
$0.97$. \label{fig:fdmba0}}
\end{figure}

For comparison, we devise another method for generating \lyf{}s from 
hydrodynamical simulations, in which the temperature of the gas particle
is calculated assuming that the baryons are in thermal equilibrium, i.e.
we discard the actual particle temperature from the simulations. 
We refer to this method as BA-TE-IE, and the name HYDRO is given 
to the full-hydrodynamical approach described in Section 
\ref{sec:hyd-lya}. One can assess the importance of shocked gas by 
comparing the method HYDRO with BA-TE-IE, while the difference between 
methods BA-TE-IE and 
DM-TE-IE must arise from differences in the baryon and
dark matter distributions. The four methods are summarized in Table 
\ref{tab:psh}.

\subsection{Comparison}
\begin{figure}
\centering
\includegraphics[width=84mm]{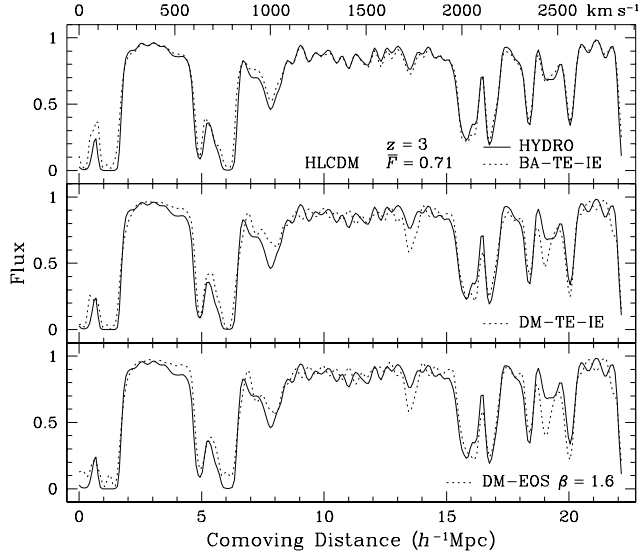}
\caption{The same as Fig.~\ref{fig:fdmba0}, except 
that the \lyf{}s are generated from baryon and dark matter 
distributions at $z = 3$ and the mean flux is $0.71$.
\label{fig:fdmba3}}
\end{figure}

To give a visual impression of pseudo-hydro techniques, 
in Figures~\ref{fig:fdmba0} and \ref{fig:fdmba3} we present \lyf{}s 
obtained along the same LOS using the four methods,
HYDRO, BA-TE-IE, DM-TE-IE, and DM-EOS. We require that 
the mean flux over all $256^2$ \lyf{}s
in all four methods match the mean flux of 
$0.97$ at $z = 0$ and $0.71$ at $z = 3$, but the mean flux of a single 
LOS is not necessarily the same across the methods. Since neither 
a simple EOS nor thermal equilibrium takes into account the 
substantial amount of warm-hot 
intergalactic medium \citep[WHIM,][]{dhk99, dt01, dco01} at $z = 0$,
 pseudo-hydro techniques are expected to be less accurate at lower 
redshift. This is seen in Figure~\ref{fig:fdmba0}. Conversely, at 
$z = 3$ methods HYDRO and BA-TE-IE generate nearly identical \lyf{}s, and 
the difference between the Ly$\alpha$ 
forests generated from baryons and those from dark matter is also small.

\begin{figure}
\centering
\includegraphics[width=84mm]{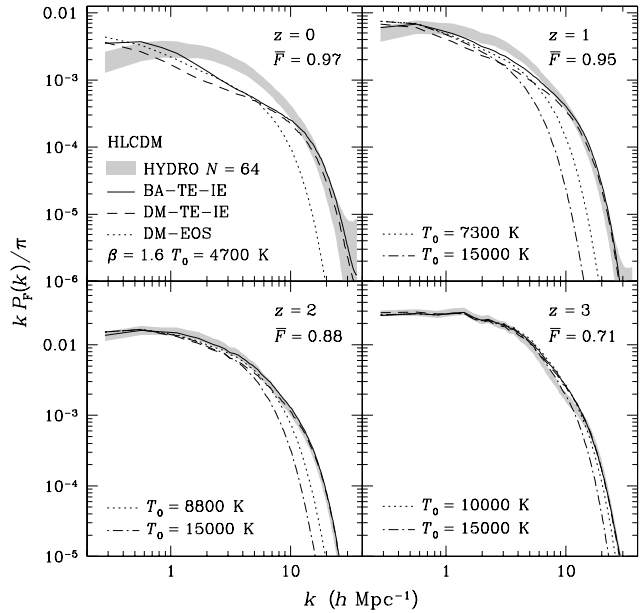}
\caption{Flux PSs of \lyf{}s at $z = 0$, 1, 2, and 3. The \lyf{}s
are produced using the four methods listed in Table \ref{tab:psh}. Grey 
bands represent standard deviations of flux PSs of \lyf{}s 
generated using the method HYDRO. The standard deviations are
calculated among 1000 groups, each of which consists of 64 LOSs. 
Additional flux PSs (dash-dotted lines) are calculated using the method 
DM-EOS with $T_0 = 15000$ K for $z = 1$, 2, and 3. Note that the flux
PSs are plotted in dimensionless form, i.e. $k P_{\rm F}(k) / \pi$.
\label{fig:fpshion}}
\end{figure}

\begin{figure}
\centering
\includegraphics[width=84mm]{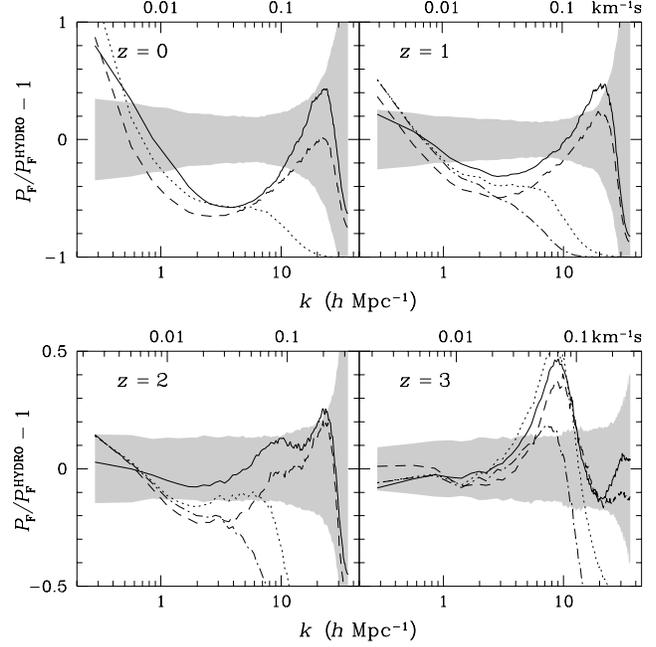}
\caption{Fractional errors in the flux PS. The legends are the 
same as in Fig.~\ref{fig:fpshion}.
\label{fig:fpsfrac}}
\end{figure}

Figure~\ref{fig:fpshion} evaluates the statistical performance of 
pseudo-hydro techniques using flux PSs, and the fractional 
errors relative to the method HYDRO are shown in 
Figure~\ref{fig:fpsfrac}. The grey bands are the standard 
deviations of the flux PSs for \lyf{}s produced using the method HYDRO. 
The standard deviations are calculated among 1000 groups, each of 
which contains 
64 LOSs randomly selected with no repetition. The total length
of 64 LOSs is 180000 \mbox{km~s$^{-1}$} at $z = 3$, about 10\% less 
than the corresponding low resolution sample in \citet{cwb02}.
There is a good agreement among all methods at $z = 3$ for
$k$ less than a few \mpci{} with a less-than-10\% difference in 
the slopes of the flux PS. Whereas, all the three pseudo-hydro methods, 
BA-TE-IE (solid lines), DM-TE-IE (dashed lines), and 
DM-EOS (dotted lines) fail to converge on HYDRO (grey bands) at $z = 0$
owing to the WHIM. In fact, methods HYDRO and BA-TE-IE have 
identical baryon distributions, so that the difference in their flux 
PSs can only be attributed to the IGM temperature, which is 
greatly affected by shock heating at low redshift. Hence, one can 
conclude that the temperature structure of the IGM is critical to the 
low-redshift \lyf{} and flux PS. Furthermore, the sensitivity of
the low-redshift \lyf{} to hydrodynamical effects forewarns us of the 
importance of other astrophysical effects, which could make the 
low-redshift \lyf{} an ideal test for realistic hydrodynamical 
simulations.

The mean-density temperature of the IGM, $T_0$, does not alter the 
optical depth in the method DM-EOS because it is absorbed into the 
constant $A$ in the approximation 
$F= \exp\left[-A (\rho/\bar{\rho})^\beta\right]$, which 
is adjusted to fit the observed mean flux. However, $T_0$ can affect 
simulated \lyf{}s through thermal broadening as indicated by
the fast drop at large $k$ of the flux PSs for the method DM-EOS. 
To test this, 
We reproduce \lyf{}s at $z = 1$, 2, and 3 using $T_0 = 15000$ K, which 
is 1.5 to 2 times the mean-density temperature of the IGM in HLCDM. 
Flux PSs of these \lyf{}s are shown in dash-dotted lines in
Figures~\ref{fig:fpshion} and \ref{fig:fpsfrac}. One sees that the higher
mean-density temperature reduces more flux power on small scales while 
leaving flux PSs unchanged on large scales. 

The poor performance of the pseudo-hydro techniques at low 
redshift means that there will be a substantial systematic error in the 
recovered \oned{} linear mass PS if it is obtained by applying the 
pseudo-hydro ratio between the \oned{} linear mass PS and flux PS 
to the observed low-redshift flux PS. For example, at $z = 2$ the slope 
of the flux PS of the method DM-EOS differ from that of HYDRO by 32\% to 
-19\% (rms 17\%) within $0.3\ \mpci < k < 2\ \mpci$, yet both flux PSs 
have the same underlying \oned{} linear mass PS. Thus, the DM-EOS method
will recover a \oned{} mass PS that is 32\% to -19\% off compared to HYDRO.
Since the slope of the \oned{} mass PS determines the shape of the 
\thrd{} mass PS (the amplitude has to be calibrated separately), the error 
in the slope will give rise to an error in the shape of the recovered 
\thrd{} linear mass PS.

\begin{figure}
\centering
\includegraphics[width=84mm]{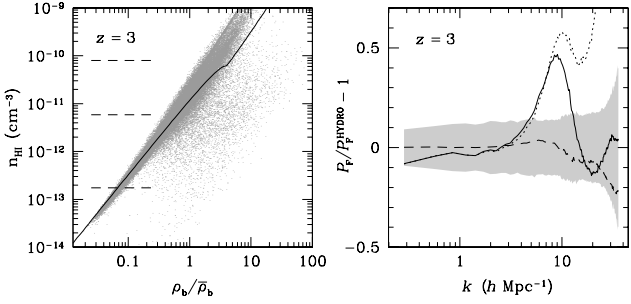}
\caption{{\it Left panel}: The neutral hydrogen number density 
$n_{\rm HI}$ as a function of the baryon density contrast 
$\rho_{\rm b}/\bar{\rho}_{\rm b}$. The grey dots correspond to the 
values of $2.5\%$ gas particles in the simulation HLCDM that are 
calculated using the method HYDRO, and  
the solid line is given by thermal equilibrium and
ionization equilibrium, i.e.~the method BA-TE-IE. From top to bottom, 
the three horizontal dashed lines mark the H{\sc i} number densities 
that would give rise to a pixel flux level of 0.01, 0.71 (the mean 
flux at $z = 3$), and 0.99, if the H{\sc i} number densities were
H{\sc i} number densities of the node.
{\it Right panel}: Fractional error in the flux PS. The grey band 
represents the standard deviation of the flux PS of \lyf{}s 
generated using the method HYDRO as in Fig.~\ref{fig:fpshion}, 
and the lines are flux PSs
of \lyf{}s generated using the method BA-TE-IE (solid line),  
BA-TE-IE for $\rho_{\rm b}/\bar{\rho}_{\rm b} > 1$ and HYDRO for
$\rho_{\rm b}/\bar{\rho}_{\rm b} \leq 1$ (dashed line), and
HYDRO for $\rho_{\rm b}/\bar{\rho}_{\rm b} > 1$ and BA-TE-IE for
$\rho_{\rm b}/\bar{\rho}_{\rm b} \leq 1$ (dotted line).
\label{fig:nHI}}
\end{figure}

Pseudo-hydro techniques replace the complex distribution of the 
neutral hydrogen number density $n_{\rm HI}$ at a given baryon 
density contrast $\rho_{\rm b}/\bar{\rho_{\rm b}}$ with a single 
function (see the left panel of Figure \ref{fig:nHI}). This 
approximation has two effects: 1) it decreases (increases) the optical 
depth of some particles and depresses (amplifies) flux fluctuations, 
and 2) it reduces the scatter in the optical depth at a given density 
and smoothes the flux. The former is more important at low densities 
where pseudo-hydro techniques tend to overestimate the optical depth, 
while the latter is more important at high densities. This is 
supported by Figure \ref{fig:fdmba3}, in which pseudo-hydro fluxes 
display richer structures and deeper absorptions than the full-hydro 
flux in regions above the mean flux.

\begin{figure}
\centering
\includegraphics[width=82mm]{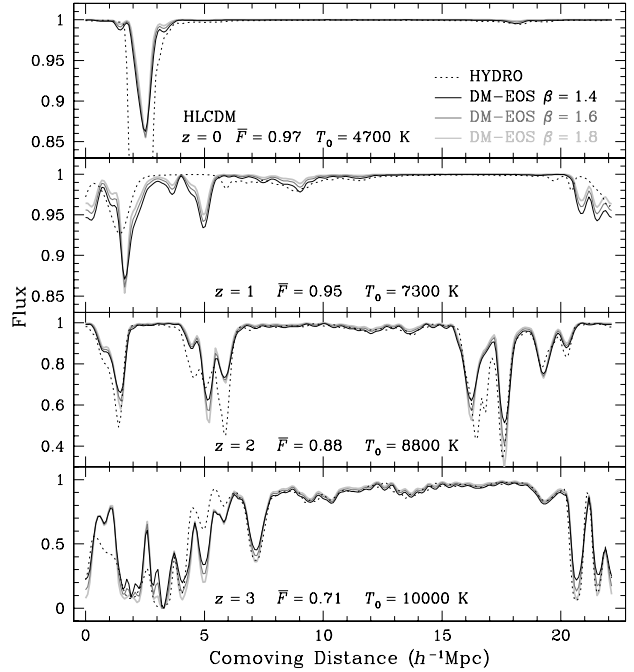}
\caption{\lyf{}s generated from dark matter distributions with 
different EOSs. The \lyf{}s are produced from dark matter densities 
using $F = \exp[-A(\rho/\bar{\rho})^\beta]$, where $A$ is adjusted to 
fit the mean flux $\bar{F}$. Thermal broadening is included with 
temperature given by $T = T_0(\rho/\bar{\rho})^{\gamma-1}$.
\label{fig:feos}}
\end{figure}

Around $k = 10$ \mpci{}, there is a relative increase of the flux PS 
from methods BA-TE-IE and DM-TE-IE with respect to that from HYDRO 
in all the redshift panels in Figure \ref{fig:fpsfrac}. This appears 
to be
the result of the competition between the increase of the optical depth 
at low densities and the reduction of the scatter of the optical depth 
at high densities in the pseudo-hydro methods. The flux PS from the 
method DM-EOS shares the same 
characteristics at $z = 3$, but stronger thermal broadening at lower 
redshift wipes out the irregularity on small scales.

To try to isolate whether it is the low or high density regions that
are responsible for the discrepancies in the pseudo-hydro flux PS,
we consider what happens when one applies pseudo-hydro to one density
regime and full hydro the rest.
We show in the right panel of Figure \ref{fig:nHI} 
the fractional error in the flux PS obtained by applying the method 
HYDRO to $\rho_{\rm b}/\bar{\rho_{\rm b}} \leq 1$ particles and BA-TE-IE 
to $\rho_{\rm b}/\bar{\rho_{\rm b}} > 1$ particles (dashed line). With
such a combination, the small-scale flux fluctuations are suppressed
because full hydro produces shallower absorptions in low density 
regions and because pseudo-hydro eliminates those fluctuations that 
arise from the scatter of the optical depth in high density regions.
Conversely, the opposite combination (dotted line) results in a boost 
of the flux PS on small scales.
It is also interesting to note that the large-scale flux PS 
($k < 2$ \mpci{}) is determined by the method that 
is applied to the low-density particles.
Hence, Figure \ref{fig:nHI} suggests that the low-density particles and 
their treatment carry a considerable weight in the Ly$\alpha$ flux PS 
on all scales.

\subsection{Tuning the Equation of State}
The equation of state (EOS)
maps density fluctuations to flux fluctuations by relating 
optical depths to densities. For a given density and mean flux, 
different EOSs will 
assign different optical depths, which will then alter the amplitude of the 
flux fluctuations and, therefore, the flux PS. 

\begin{figure}
\centering
\includegraphics[width=82mm]{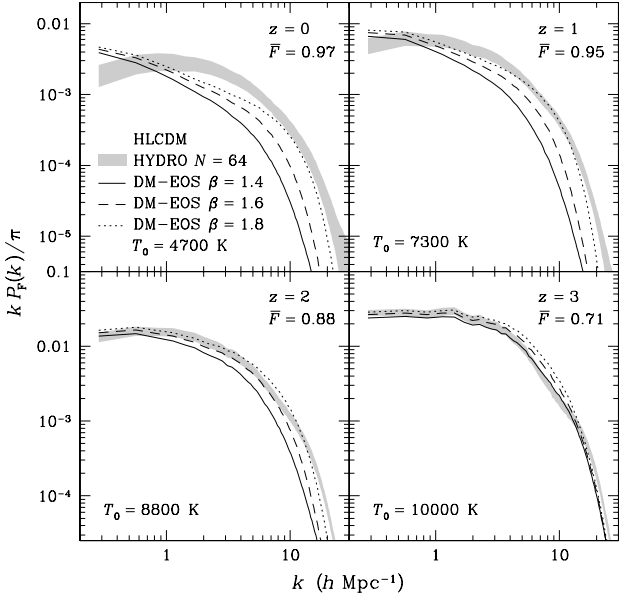}
\caption{Flux PSs of \lyf{}s generated from dark-matter-converted 
baryon densities using different EOSs. Grey bands represent 
standard deviations of the flux PSs of \lyf{}s generated using the
method HYDRO as in Fig.~\ref{fig:fpshion}. Also as in Fig.~\ref{fig:fpshion}
dimensionless PSs are plotted.
\label{fig:fpseos}}
\end{figure}

For a stiffer EOS, i.e.~a smaller value of $\beta$ (larger $\gamma$), 
high-density regions 
have to absorb less Ly$\alpha$ flux, while, in compensation, low-density 
regions have to absorb more flux. In terms of flux, a stiff EOS leads to higher 
fluxes in deep (or large-equivalent-width) absorptions and lower fluxes in 
shallow absorptions than a soft EOS. This expectation is confirmed in 
Figure~\ref{fig:feos}, where \lyf{}s generated using the method DM-EOS are 
compared with those using the method HYDRO at 
$z = 0$, 1, 2, and 3. The mean flux is kept the same for both methods 
at each epoch, varying only the EOS. The value of 
$\beta = 1.4$ in the figures corresponds to a very stiff EOS, 
i.e.~$\gamma = 1.86$, and it is provided only for the purpose of 
comparison.

Figure~\ref{fig:feos} shows that low-amplitude and 
small-scale fluctuations in the flux are likely to be suppressed by the
method DM-EOS. This reduces the flux PS on small scales as seen in 
Figures~\ref{fig:fpshion} and \ref{fig:fpsfrac}. The method DM-EOS is not a 
good approximation at low redshift, but it improves as redshift increases.

Since the amplitude of flux fluctuations increases with $\beta$ in 
Figure~\ref{fig:feos}, a smaller value of $\beta$
must lead to a lower flux PS. This is observed in Figure~\ref{fig:fpseos}, 
where flux PSs of \lyf{}s obtained using the method DM-EOS with different
EOSs are compared with those using the method HYDRO. 
Figure~\ref{fig:fpseos} demonstrates that one cannot tune the EOS to make 
the pseudo-hydro method DM-EOS work at low redshift. Again, the method 
DM-EOS appears to be a reasonable approximation for studies of the flux PS at 
$z = 3$, although it may not be true for higher-order statistics. 
The difference among different EOSs is also less pronounced at $z = 3$ 
because the dynamic range of the density contrast,
$\rho / \bar{\rho}$, is much smaller. 

\section{Mass Statistics vs. Flux Statistics} \label{sec:comp}
\subsection{Power Spectrum}
The \lyf{} has been used to infer the linear mass PS of the cosmic
density field. The nonlinear transform of the density to the flux has 
made it difficult to derive the (linear) mass PS from the flux PS 
analytically. One way to circumvent this difficulty is to use simulations 
to map the flux PS to the linear mass PS \citep[e.g.][]{cwb02}. 
Although the flux PS resembles the linear \oned{} mass PS, the
underlying nonlinear density field is what produces the \lyf{}.
As such, it is important to compare 
the flux PS with the mass PS. 

\begin{figure}
\centering
\includegraphics[width=84mm]{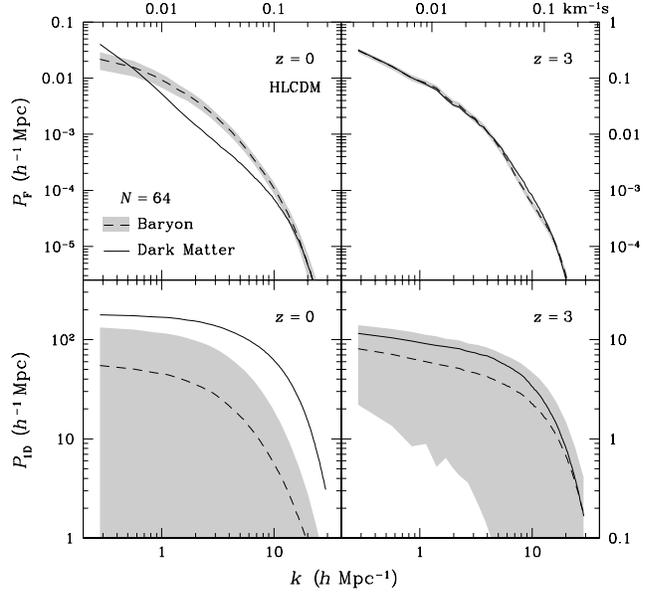}
\caption{Flux PSs of simulated \lyf{}s and one-dimensional mass PSs 
of the underlying density fields at $z = 0$ and 3. The \lyf{}s are 
produced with methods HYDRO and DM-TE-IE from baryon and dark matter
distributions, respectively.
\label{fig:mfps03}}
\end{figure}

Plotted in Figure~\ref{fig:mfps03} are flux PSs 
produced using methods HYDRO and DM-TE-IE
along with one-dimensional mass PSs of baryons and dark matter. 
The standard deviations of baryon flux PSs and mass PSs are shown in 
grey bands. The most prominent feature is that one-dimensional mass 
PSs have much larger dispersions than flux PSs. As discussed in 
\citet{ze05}, the variance in the one-dimensional mass PS is 
severely inflated by the trispectrum of the cosmic density field because 
of the non-Gaussianity. 

An interesting observation is that unlike the mass PS the flux PS 
decreases with time. This is due to the thinning of the \lyf{} and the 
higher mean flux toward lower redshift that reduce fluctuations in the 
Ly$\alpha$ flux.

The nonlinear transform between baryon density and 
flux greatly suppresses the fluctuations. For example, the overdensity 
$\delta$ may vary from -1 to hundreds (tens) 
at $z = 0$ ($z = 3$), but the flux can only be between 0 and 1. 
With a mean flux on the order of unity, fluctuations in the flux are 
$10$ to $10^2$ times smaller than those in the cosmic density field. 
Hence, the flux PS is a factor of $10^2$ ($z = 3$) to $10^4$ ($z = 0$) 
times lower than the one-dimensional mass PS. Moreover, the 
non-Gaussianity in the cosmic density field is also strongly suppressed 
in the flux. 
Thus, the flux trispectrum is much closer to zero as compared to the mass 
trispectrum of the cosmic density field, and the variance of the flux PS 
becomes much smaller than the variance of the one-dimensional mass PS.

The near-Gaussian Ly$\alpha$ flux is probably the reason that many 
simulations and techniques are able to reproduce lower-order statistics 
of the observed \lyf{}, especially at high redshift. 
Figure~\ref{fig:mfps03} points out a possible problem that can arise:
one could produce \lyf{}s from wildly different density 
fields but still have almost identical flux PSs. For example, even though 
baryons and dark matter differ considerably in terms of mass PS 
(see also Figures~\ref{fig:cosmfps0} and \ref{fig:cosmfps3}), 
they are not so distinguished from each 
other in flux PSs at $z \geq 2$. Conversely, we are able to measure 
the flux PS extremely well, but the underlying mass PS may still be much
less constrained.

\subsection{Covariance}
The covariance of the PS is a fourth-order statistic that measures the 
uncertainties in the PS as well as the correlation between modes.
Here, we use it to explore the difference 
between Ly$\alpha$ forests generated using the full-hydro method HYDRO 
and those using the pseudo-hydro method DM-TE-IE.

\begin{figure}
\centering
\includegraphics[width=84mm]{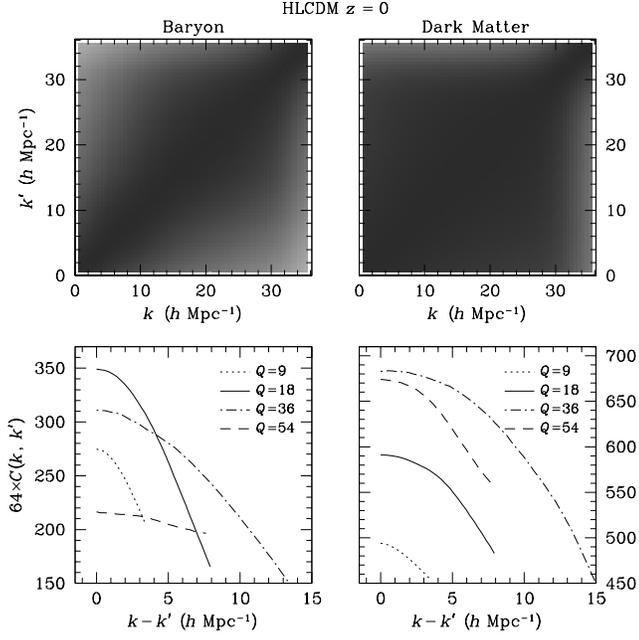}
\caption{Covariances of one-dimensional baryon and dark matter 
PSs at $z = 0$. The upper panels are reduced covariances
$\hat{C}(k, k')$ in a linear grey scale with black being $1.2$ and 
white less than or equal to 0.
The lower panels are cross sections of normalized covariances
$C(k, k')$ along $Q = (k + k') / \mpc$. The covariances 
$C(k, k')$ are multiplied by 64 for better comparison with that of
GRFs, which follows $64\,C(k, k') = \delta^{\rm K}_{n, n'}$.
All the covariances are calculated from 1000 groups, each 
of which consists of 64 LOSs ($N=64$) randomly selected from the density 
grid of HLCDM. 
\label{fig:cv0db64}}
\end{figure}

\begin{figure}
\centering
\includegraphics[width=84mm]{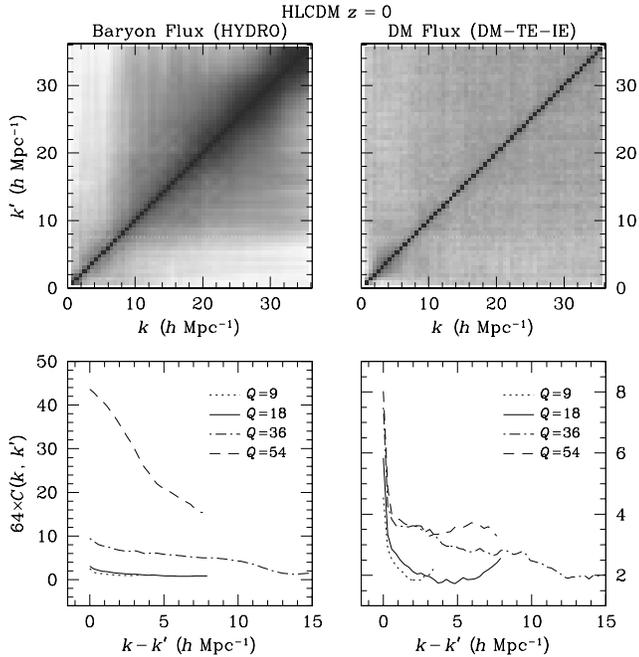}
\caption{The same as Fig.~\ref{fig:cv0db64}, but for flux PSs of \lyf{}s
generated from baryons and dark matter using methods HYDRO and DM-TE-IE,
respectively.
\label{fig:cv0fdb64}}
\end{figure}

\begin{figure}
\centering
\includegraphics[width=84mm]{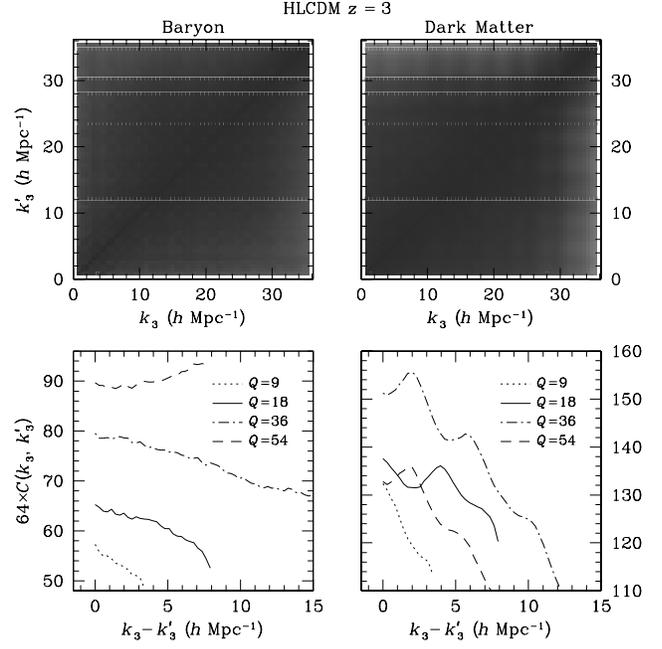}
\caption{The same as Fig.~\ref{fig:cv0db64}, but for the simulation 
HLCDM at $z = 3$.
\label{fig:cv3db64w}}
\end{figure}

\begin{figure}
\centering
\includegraphics[width=84mm]{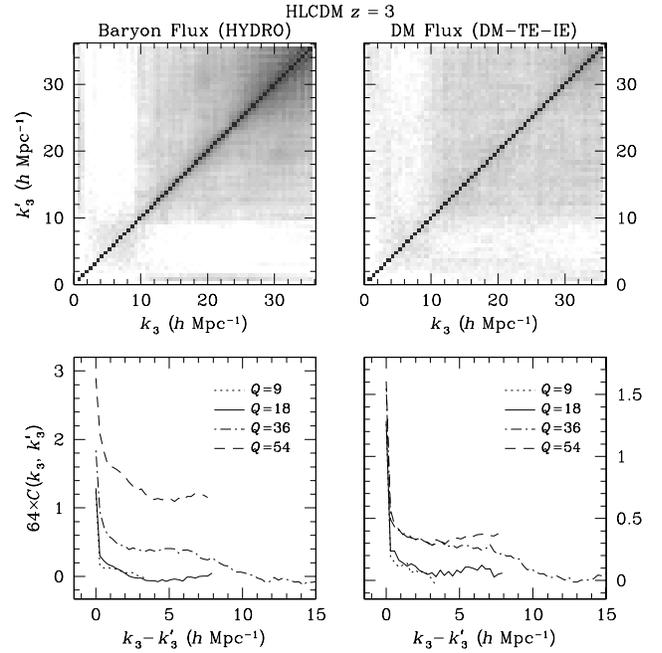}
\caption{The same as Fig.~\ref{fig:cv3db64w}, but for flux PSs.
\label{fig:cv3fdb64w}}
\end{figure}

The covariance of the one-dimensional mass PS is defined as
\begin{equation} \label{eq:cov-def}
\sigma^2_{\rm 1D}(k, k') = 
\langle [P_{\rm 1D}(k) - \langle P_{\rm 1D}(k) \rangle]
[P_{\rm 1D}(k') - \langle P_{\rm 1D}(k') \rangle] \rangle,
\end{equation}
where $\langle\ldots\rangle$ stands for an ensemble average, and 
$P_{\rm 1D}(k)$ can be replaced by $P_{\rm F}(k)$ for the flux PS
covariance $\sigma^2_{\rm F}(k, k')$. Since we 
use a discrete Fourier transform for the analysis, the wavenumber $k$
is discrete, i.e.~$k = 2n\pi / L$, where $n$ assumes integer values and 
$L$ is the length of the spectrum. We use $k$ and $n$ interchangeably.

There is a hidden variable $N$, the number of LOSs that are averaged 
over to obtain the one-dimensional PS, in equation (\ref{eq:cov-def}). 
For Gaussian random fields (GRFs), one can show that
\begin{equation} \label{eq:gascov}
\sigma^2_{\rm 1D}(k, k') \simeq \frac{1}{N} P_{\rm 1D}^2(k) \,
\delta^{\rm K}_{n, n'},
\end{equation}
where $\delta^{\rm K}_{n, n'}$ is the Kronecker delta function.
This also applies to the flux PS covariance. Equation 
(\ref{eq:gascov}) is only approximate owing to both the finite length of the
LOSs and the correlation between the LOSs \citep{ze05}.

We also introduce the normalized covariance
\begin{equation} \label{eq:C}
C(k, k') = \sigma^2_{\rm 1D}(k, k') [P_{\rm 1D}(k)P_{\rm 1D}(k')]^{-1},
\end{equation}
and the reduced covariance
\begin{equation} \label{eq:Chat}
\hat{C}(k, k') = C(k, k') [C(k, k)C(k', k')]^{-1/2}. 
\end{equation}
Again, $P_{\rm 1D}(k)$ can be replaced by $P_{\rm F}(k)$ for the flux 
PS covariance. For GRFs, both the covariance matrices are diagonal. In
addition, we have $N C(k, k) = 1$. The advantage of 
$\hat{C}(k, k')$ is that $\hat{C}(k, k) = 1$ for all fields, so 
that they can be compared with each other in a single grey scale.

Figures~\ref{fig:cv0db64} and \ref{fig:cv0fdb64} illustrate the 
covariances $\hat{C}(k, k')$ and $C(k, k')$ of one-dimensional mass 
PSs and flux PSs at $z = 0$. The covariances are 
calculated from 1000 groups, each of which consists of 64 LOSs ($N=64$) 
randomly selected from the density grid of HLCDM. For GRFs, the 
covariance matrix $\hat{C}(k, k')$ is diagonal, and the normalized 
variance $C(k, k)$ equals $N^{-1}$. For better comparison, the 
covariances $C(k, k')$ are multiplied by $N$, so that the
Gaussian case has $N\,C(k, k') = \delta^{\rm K}_{n, n'}$.
As already seen in \citet{ze05}, the covariances of 
one-dimensional mass PSs are starkly non-Gaussian. The variances 
in the \oned{} mass PS are two orders of magnitude higher than 
expected for GRFs.
The covariances of baryons are roughly a factor of 2 lower
than those of dark matter. This is likely due to the pressure 
experienced by the SPH particles.
The covariances of flux PSs have a dominant diagonal, though they are
still not Gaussian. 
The method HYDRO gives rise to stronger 
correlations between high-$k$ modes in the flux PS than the method DM-TE-IE (as
well as BA-TE-IE, which is not shown) because the simple EOS (or thermal 
equilibrium for BA-TE-IE) is not sufficient to describe the WHIM.

\begin{figure}
\centering
\includegraphics[width=84mm]{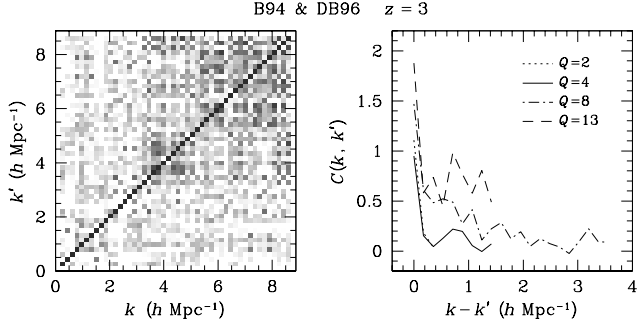}
\caption{The same as Fig.~\ref{fig:cv3fdb64w}, but for observed \lyf{}s. 
The covariances are calculated from 27 segments of \lyf{}s, i.e.~27 
groups with $N=1$. Note that the resolution of the observed \lyf{}s
is about four times lower than that in the simulations.
\label{fig:cv3fobs}}
\end{figure}

Figures~\ref{fig:cv3db64w} and \ref{fig:cv3fdb64w} present the covariances 
of one-dimensional mass PSs and flux PSs at $z = 3$. At this redshift, the 
covariances of one-dimensional mass PSs are reduced by a factor of a few, 
but they are still 
highly non-Gaussian. Whereas, the covariances of flux PSs are very close to 
Gaussian. The difference in the covariances between the two methods HYDRO and
DM-TE-IE is significantly reduced compared to that at $z = 0$. 

In addition to simulations, we show in Figure~\ref{fig:cv3fobs} the
covariances of observed flux PSs at $z = 3$. The sample of Ly$\alpha$ 
forests includes 20 QSO spectra from \citet{b94} and 
\citet{db96}. The QSO spectra 
are selected so that each contains at least one good chunk of spectrum 
that has no bad pixels or strong metal lines and spans 64 \AA{} anywhere
within $z = 2.9$--$3.1$. The spectral resolution is 1 \AA{}, which is about
four times lower than that in the simulations. In all, there are 27 
segments of \lyf{}s for analysis. We do not re-group the segments, 
i.e.~$N = 1$, in calculating the covariances.

The main characteristics of the observed covariances are in good 
agreement with simulated ones. Namely, the covariance matrices 
have a strongly dominant diagonal, and they are very close to Gaussian. The 
values of diagonal elements roughly match those in the simulations but 
the off-diagonal elements are noisier owing to the small 
sample size. With a large number of high-resolution \lyf{}s, one
will be able to study the behavior of the covariance on smaller scales
(larger $k$) and reduce statistical uncertainties.

A general observation of the covariances of flux PSs is that the 
correlation between two LOS modes decreases away from the diagonal, 
because two neighboring modes are more likely to be correlated than two 
distant modes. In most cases the correlation between modes and 
the variance of the PS increase toward small scales, over which the 
underlying density field is also more nonlinear and non-Gaussian.
Beyond these points, however, the behavior of the 
covariances is not quantitatively understood. 

\section{Cosmology} \label{sec:cos}
Because of the difficulty in deriving density statistics from flux 
statistics, one often resorts to numerical simulations and constrains 
cosmology by comparing observed flux statistics directly to simulated 
flux statistics. In addition, one utilizes fast $N$-body simulations 
and pseudo-hydro techniques in order to explore a large cosmological 
parameter space in manageable time. This necessitates an examination of
the accuracy of pseudo-hydro techniques and the sensitivity of 
flux statistics to cosmology.

\begin{figure}
\centering
\includegraphics[width=84mm]{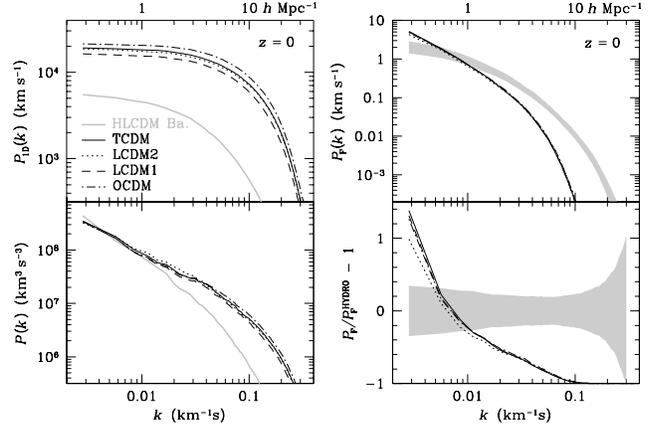}
\caption{Mass PSs of baryons and dark matter and flux PSs of 
simulated \lyf{}s for five cosmological models at $z = 0$. The 
upper left panel shows one-dimensional mass PSs, and the lower panel 
\thrd{} mass PSs. The upper right panel shows flux PSs, and the lower 
panel residuals of flux PSs with respect to the flux PS of Ly$\alpha$ 
forests generated using the method HYDRO from HLCDM (light
grey lines and bands). All other flux PSs are from $N$-body simulations 
using the method DM-EOS. 
\label{fig:cosmfps0}}
\end{figure}

\begin{figure}
\centering
\includegraphics[width=84mm]{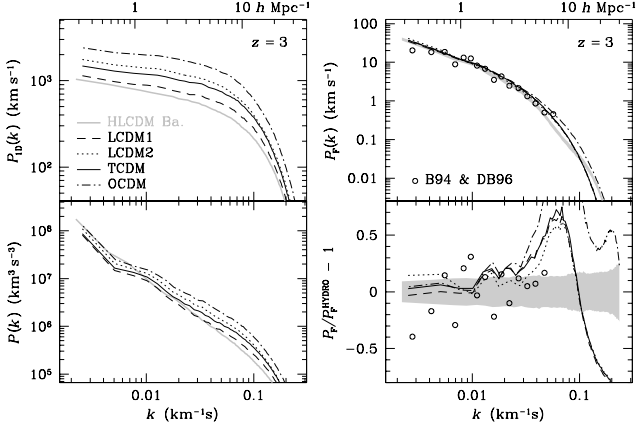}
\caption{The same as Fig.~\ref{fig:cosmfps0} except that 
all PSs are at $z = 3$. The observed flux PS is measured from 27
segments of \lyf{}s between $z = 2.9$ and $3.1$. The uncertainties of the
observed flux PS can be inferred from Figure~\ref{fig:cv3fobs}.
The wavenumbers labelled on the top are only for the HLCDM model.
\label{fig:cosmfps3}}
\end{figure}

Figures~\ref{fig:cosmfps0} and \ref{fig:cosmfps3} compare mass PSs and 
flux PSs for five simulations: HLCDM, LCDM1, high 
normalization LCDM (LCDM2), tilted LCDM (TCDM), and open CDM (OCDM). Table 
\ref{tab:para1} lists the parameters for all these models. The $N$-body 
simulations all have the same box size of $22.222$ \mpc{} on each side and
evolve $128^3$ CDM particles from $z = 49$ to 0 using {\sc gadget}
\citep{syw01}. The initial conditions of these simulations have the 
same Fourier phases. Note that 
the TCDM model has an opposite tilt than HLCDM. Not all the
simulations are consistent with most recent observations and they are 
provided only for testing the cosmological dependence of the flux PS.

Pseudo-hydro techniques have already been proven inaccurate at low redshift 
by several tests above. We include the results at $z = 0$ here only to show 
that all the flux PSs based on the method DM-EOS are nearly 
indistinguishable from each other except the high normalization model LCDM2. 

At $z = 3$, the flux PS of the OCDM model departs from others at 
$k \ga 3$ \mpci{}. However, this difference is not much more 
pronounced than those between the flux PSs obtained from the same 
simulation (HLCDM) but using different methods (see 
Figure~\ref{fig:fpsfrac}). Moreover, if the flux PS is simply a 
linearized one-dimensional mass PS, one may expect that the order of the 
mass PS amplitudes should be preserved in the flux PS, but this is not 
observed in Figure~\ref{fig:cosmfps3}. For example, the initial
mass PS of LCDM2 (dotted lines) has the same shape as that of LCDM1 
(dashed lines) but with a 56\% higher normalization. This relative 
amplitude is preserved in the nonlinear three-dimensional PS and \oned{}
PS. However, as seen in the lower right panel, the flux PS of LCDM2 
declines from 20\% higher to 10\% lower than that of LCDM1 from 
$k = 0.002$ to $0.04$~\mbox{km$^{-1}$s}. This result is roughly 
consistent with that reported by \citet{msc04b}, who see a decline from 
6\% to -1\% in the flux PS over the same range if the initial mass PS 
is boosted by 10\%.

Without the detailed knowledge of the state of the IGM, one may confuse 
the differences between cosmological models (Figure~\ref{fig:cosmfps3}) 
with the systematic errors of the pseudo-hydro methods 
(Figure~\ref{fig:fpsfrac}). Therefore, to do precision cosmology with 
the \lyf{} without overly relying on hydrodynamical simulations, one 
should, at least, have precise calibrations of pseudo-hydro techniques.

\section{Conclusions} \label{sec:con}
Using hydrodynamical simulations and $N$-body simulations we find 
that pseudo-hydro techniques are able to reproduce the flux and flux PS 
at $k$ greater than a few \mpci{} that are obtained using the 
full-hydro method at $z = 3$. There is also a good match between 
observed and simulated flux statistics such as the flux PS and its 
covariance at $z =3$. Since the performance of pseudo-hydro techniques
improves toward higher redshift, we expect them to work even better at
$z > 3$.

For pseudo-hydro techniques, the mean-density temperature of the 
IGM does not affect the mean flux of the \lyf{} but it does alter the 
flux PS on small scales ($k \ga$ several \mpci{}) through 
thermal broadening. The EOS of the IGM modifies both the amplitude and 
shape of the flux PS. We also observe a significant 
evolution in the shape of the flux PS from $z = 0$ to 3 
(e.g. Fig.~\ref{fig:fpseos}).

The accuracy of pseudo-hydro techniques does not seem to be high 
enough for determining the Ly$\alpha$ flux PS at a few percent level.
One needs to precisely calibrate pseudo-hydro techniques with
hydrodynamical simulations. Moreover, it is better to constrain 
cosmology using the flux PS on scales above a few \mpc{} to 
reduce the uncertainties caused by the incomplete knowledge of the 
IGM. 

To infer the one-dimensional linear mass PS one often divides 
the observed flux PS by the ratio between the simulated 
three-dimensional flux PS and the theoretical linear mass PS. This
procedure is widely tested \citep[e.g.][]{gh02}.
In this way the sample variance error in the mass PS of the underly 
density field is avoided.

The transform from density to flux quenches fluctuations by orders of 
magnitude and leads to near-Gaussian Ly$\alpha$ fluxes. Hence, the 
variance of the flux PS is much less than that of the one-dimensional 
mass PS. In other words, one can measure the flux 
PS to a high precision with a relatively small number of LOSs but the
underlying mass PS cannot be determined as precisely as the flux PS.
Therefore, a large number of LOSs are needed to reduce the sample 
variance error in the one-dimensional mass PS, 
e.g. without binning the modes the standard deviation of the 
mean mass PS of 1024 LOSs is roughly 17\% in a 
simulated cosmic density field at $z = 3$ \citep{ze05}. Since the
sample variance error in the one-dimensional mass PS is approximately 
inversely proportional to the number of LOSs, attempts to recover the 
\thrd{} mass PS accurate to 5\% may require more than 10,000 LOSs. 
Such a large number of LOSs is very demanding, but 
is still within the reach of the Sloan Digital Sky Survey (SDSS).
For instance, \citet{msc04b} have used more than 3000 
$z > 2.3$ quasar spectra from the SDSS (over roughly a quarter of the
targeted sky coverage) to estimate the linear mass PS. Thus, it is 
reasonable to project that there will be more than 10000 quasar
spectra available when the SDSS finishes. Conversely, one can 
also trade resolution with precision by binning the modes and mitigate 
the demand for LOSs, even though binning highly correlated Fourier 
modes in the density field is not as efficient in reducing the sample 
variance error as binning independent ones.

The growing nonlinearity and non-Gaussianity at lower
redshift drives up the cosmic variance of the mass PS and the
correlation between different modes in the density field. This means 
that even more LOSs are needed to extend the Ly$\alpha$ flux PS 
analysis to low redshift. Fortunately, there are far more low-redshift
quasar spectra available than high-redshift ones, although there will 
be many other astrophysical and observational challenges to be 
addressed. One such challenge is the inadequacy of the pseudo-hydro 
techniques at low redshift (Figure~\ref{fig:fpshion}), when the 
shock-heated WHIM greatly alters the temperature--density relation of 
the IGM. A recent comparison between the HPM and full hydro methods 
confirms that it is the hot-phase gas that causes the differences 
between full and pseudo hydro results \citep{vhs05}. 
Thus it seems inevitable that 
time-consuming hydrodynamical simulations are needed to accurately 
model the low-redshift \lyf{} and to provide the means for inferring 
the mass PS and cosmological parameters.

\section*{Acknowledgments}
We thank D.~Weinberg and the referee for helpful comments.
HZ was supported by the NSF under grant 0307961 and NASA under grant
NAG5-11098.
DJE was supported by NSF AST-0098577 and an Alfred P.~Sloan Research 
Fellowship. NSK was supported by NSF AST-0205969 and NASA NAGS-13308.


\begin{thebibliography}{}

\bibitem[\protect\citeauthoryear{Bechtold}{1994}]{b94} Bechtold J., 
1994, ApJS, 91, 1

\bibitem[\protect\citeauthoryear{Bi \& Davidsen}{1997}]{bd97} Bi H.~G.,
Davidsen A.~F., 1997, ApJ, 479, 523 


\bibitem[\protect\citeauthoryear{Cen et al.}{1994}]{cmo94} Cen R.,
Miralda-Escud\'e J., Ostriker J.~P., Rauch M., 1994, ApJ, 437, L9

\bibitem[\protect\citeauthoryear{Croft et al.}{1997}]{cwk97} 
Croft R.~A.~C., Weinberg D.~H., Katz N., Hernquist L., 1997, ApJ, 
488, 532

\bibitem[\protect\citeauthoryear{Croft et al.}{1998}]{cwk98} 
Croft R.~A.~C., Weinberg D.~H., Katz N., Hernquist L., 1998, ApJ, 
495, 44

\bibitem[\protect\citeauthoryear{Croft et al.}{1999}]{cwp99} 
Croft R.~A.~C., Weinberg D.~H., Pettini M., Hernquist L., Katz N., 
1999, ApJ, 520, 1

\bibitem[\protect\citeauthoryear{Croft et al.}{2002}]{cwb02} 
Croft R.~A.~C., Weinberg D.~H., Bolte M., Burles S., Hernquist L., 
Katz N., Kirkman D., Tytler D., 2002, ApJ, 581, 20

\bibitem[\protect\citeauthoryear{Dav\'e \& Tripp}{2001}]{dt01} 
Dav\'{e} R., Tripp T.~M., 2001, ApJ, 553, 528

\bibitem[\protect\citeauthoryear{Dav\'{e}, Dubinski \& Hernquist}{1997}]
{ddh97} Dav\'{e} R., Dubinski J., Hernquist L., 1997, New Astro., 2, 277

\bibitem[\protect\citeauthoryear{Dav\'e et al.}{1997}]{dhw97} 
Dav\'{e} R., Hernquist L., Weinberg D.~H., Katz N., 1997, ApJ, 477, 21

\bibitem[\protect\citeauthoryear{Dav\'e et al.}{1999}]{dhk99} 
Dav\'{e} R., Hernquist L., Katz N., Weinberg D.~H., 1999, ApJ, 511, 521

\bibitem[\protect\citeauthoryear{Dav\'e et al.}{2001}]{dco01} 
Dav\'e R., Cen R., Ostriker, J.~P., Bryan G.~L., Hernquist L., Katz N., 
Weinberg D.~H., Norman M.~L., O'Shea B., 2001, ApJ, 552, 473

\bibitem[\protect\citeauthoryear{Dobrzycki \& Bechtold}{1996}]{db96} 
Dobrzycki A., Bechtold J., 1996, ApJ, 457, 102

\bibitem[\protect\citeauthoryear{Gnedin \& Hamilton}{2002}]{gh02}
Gnedin N.~Y., Hamilton A.~J.~S., 2002, MNRAS, 334, 107

\bibitem[\protect\citeauthoryear{Gnedin \& Hui}{1998}]{gh98} 
Gnedin N.~Y., Hui L., 1998, MNRAS, 296, 44

\bibitem[\protect\citeauthoryear{Haardt \& Madau}{1996}]{hm96} 
Haardt F., Madau P., 1996, ApJ, 461, 20

\bibitem[\protect\citeauthoryear{Hernquist et al.}{1996}]{hkw96} 
Hernquist L., Katz N., Weinberg D.~H., Miralda-Escud\'{e} J., 1996, 
ApJ, 457, L51

\bibitem[\protect\citeauthoryear{Hui}{1999}]{h99} Hui L., 1999, ApJ, 
516, 519

\bibitem[\protect\citeauthoryear{Hui \& Gnedin}{1997}]{hg97} Hui L., 
Gnedin N.~Y., 1997, MNRAS, 292, 27

\bibitem[\protect\citeauthoryear{Hui et al.}{2001}]{hbs01} Hui L., 
Burles S., Seljak U., Rutledge R.~E., Magnier E., Tytler D., 2001, ApJ,
552, 15



\bibitem[\protect\citeauthoryear{Katz, Weinberg \& Hernquist}{1996}]
{kwh96} Katz N., Weinberg D.~H., Hernquist L., 1996, ApJS, 105, 19

\bibitem[\protect\citeauthoryear{Kim et al.}{2002}]{kcc02} Kim T.-S., 
Carswell R.~F., Cristiani S., D'Odorico S., Giallongo E., 2002, MNRAS, 
335, 555

\bibitem[\protect\citeauthoryear{Kim et al.}{2004}]{kvh04} Kim T.-S., 
Viel M., Haehnelt M.~G., Carswell R.~F., Cristiani S., 2004, MNRAS, 
347, 355

\bibitem[\protect\citeauthoryear{Lidz et al.}{2005}]{lhh05}
Lidz A., Heitmann K., Hui L., Habib S., Rauch M., Sargent W.~L.~W.,
2005, astro-ph/0505138

\bibitem[\protect\citeauthoryear{Feng \& Fang}{2000}]{ff00} Feng L.-L., 
Fang L.-Z., 2000, ApJ, 535, 519

\bibitem[\protect\citeauthoryear{Feng, Pando \& Fang}{2001}]{fpf01} 
Feng L.-L., Pando J., Fang L.-Z., 2001, ApJ, 555, 74 


\bibitem[\protect\citeauthoryear{Lu et al.}{1996}]{lsw96} Lu L., 
Sargent W.~L.~W., Womble D.~S., Takada-Hidai M., 1996, ApJ, 472, 509


\bibitem[\protect\citeauthoryear{Mandelbaum et al.}{2003}]{mms03} 
Mandelbaum R., McDonald P., Seljak U., Cen R., 2003, MNRAS, 344, 776

\bibitem[\protect\citeauthoryear{McDonald}{2003}]{m03} McDonald P., 
2003, ApJ, 585, 34

\bibitem[\protect\citeauthoryear{McDonald \& Miralda-Escud\'{e}}{1999}]
{mm99} McDonald P., Miralda-Escud\'{e} J., 1999, ApJ, 518, 24

\bibitem[\protect\citeauthoryear{McDonald et al.}{2000}]{mmr00} 
McDonald P., Miralda-Escud\'{e} J., Rauch M., Sargent W.~L.~W., 
Barlow T.~A., Cen R., Ostriker J.~P., 2000, ApJ, 543, 1

\bibitem[\protect\citeauthoryear{McDonald et al.}{2004a}]{msc04a} 
McDonald P., Seljak U., Cen R., Bode P., Ostriker J.~P., 2004, 
submitted to MNRAS (astro-ph/0407378)

\bibitem[\protect\citeauthoryear{McDonald et al.}{2004b}]{msc04b} 
McDonald P. et al., 2004, submitted to ApJ (astro-ph/0407377)

\bibitem[\protect\citeauthoryear{Meiksin \& White}{2004}]{mw04} 
Meiksin A., White M., 2004, MNRAS, 350, 1107


\bibitem[\protect\citeauthoryear{Monaghan \& Lattanzio}{1985}]{ml85} 
Monaghan J.~J., Lattanzio J.~C., 1985, A\&{}A, 149, 135

\bibitem[\protect\citeauthoryear{Nusser \& Haehnelt}{1999}]{nh99} 
Nusser A., Haehnelt M., 1999, MNRAS, 303, 179

\bibitem[\protect\citeauthoryear{Penton, Stocke and Shull}{2004}]{pss04}
Penton S.~V., Stocke J.~T., Shull J.~M., 2004, ApJS, 152, 29

\bibitem[\protect\citeauthoryear{Petitjean, M\"ucket \& Kates}{1995}]
{pmk95} Petitjean P., M\"ucket J.~P., Kates R.~E., 1995, A\&{}A, 295, L9

\bibitem[\protect\citeauthoryear{Pichon et al.}{2001}]{pvr01} Pichon C., 
Vergely J.~L., Rollinde E., Colombi S., Petitjean P., 2001, MNRAS, 326, 
597



\bibitem[\protect\citeauthoryear{Rauch et al.}{1997}]{rms97} 
Rauch M. et al., 1997, ApJ, 489, 7

\bibitem[\protect\citeauthoryear{Riediger, Petitjean \& M\"ucket}{1998}]
{rpm98} Riediger R., Petitjean P., M\"ucket J.~P., 1998, A\&{}A, 329, 30

\bibitem[\protect\citeauthoryear{Seljak, McDonald \& Makarov}{2003}]
{smm03} Seljak U., McDonald P., Makarov A., 2003, MNRAS, 342, L79

\bibitem[\protect\citeauthoryear{Seljak et al.}{2004}]{smm04} 
Seljak U. et al., 2004, submitted to Phys Rev D (astro-ph/0407372)

\bibitem[\protect\citeauthoryear{Spergel et al.}{2003}]{svp03}
Spergel D.~N. et al., 2003, ApJS, 148, 175


\bibitem[\protect\citeauthoryear{Springel, Yoshida \& White}{2001}]
{syw01} Springel V., Yoshida N., White S.~D.~M., 2001, New Astro., 
6, 79



\bibitem[\protect\citeauthoryear{Viel et al.}{Viel, Haehnelt
\& Springel}{2005}]{vhs05}
Viel M., Haehnelt M.~G., Springel V., 2005, submitted to MNRAS
(astro-ph/0504641)

\bibitem[\protect\citeauthoryear{Viel et al.}{Viel, Weller \& 
Haehnelt}{2004}]{vwh04}
Viel M., Weller J., Haehnelt M.~G., 2004, MNRAS, 355, L23

\bibitem[\protect\citeauthoryear{Viel et al.}{2003}]{vmt03} Viel M., 
Matarrese S., Theuns T., Munshi D., Wang Y., 2003, MNRAS, 340, L47

\bibitem[\protect\citeauthoryear{Viel et al.}{2004}]{vmh04} Viel M., 
Matarrese S., Heavens A., Haehnelt M.~G., Kim T.-S., Springel V., 
Hernquist L., 2004, MNRAS, 347, L26

\bibitem[\protect\citeauthoryear{Viel et al.}{2004}]{vhc04} Viel M., 
Haehnelt M.~G., Carswel R.~F., Kim T.-S., 2004, MNRAS, 349, L33

\bibitem[\protect\citeauthoryear{Zaldarriaga, Hui \& Tegmark}{2001}]
{zht01} Zaldarriaga M., Hui L., Tegmark M., 2001, ApJ, 557, 519

\bibitem[\protect\citeauthoryear{Zaldarriaga, Scoccimarro \& Hui}{2003}]
{zsh03} Zaldarriaga M., Scoccimarro R., Hui L., 2003, ApJ, 590, 1

\bibitem[\protect\citeauthoryear{Zhan}{2003}]{z03} Zhan H., 2003, MNRAS,
344, 935


\bibitem[\protect\citeauthoryear{Zhan \& Eisenstein}{2005}]{ze05} 
Zhan H., Eisenstein D., 2005, MNRAS, 357, 1387

\bibitem[\protect\citeauthoryear{Zhan \& Fang}{2002}]{zf02} Zhan H., 
Fang L.-Z., 2002, ApJ, 566, 9


\bibitem[\protect\citeauthoryear{Zhan, Jamkhedkar \& Fang}{2001}]{zjf01}
Zhan H., Jamkhedkar P., Fang L.-Z., 2001, ApJ, 555, 58

\bibitem[\protect\citeauthoryear{Zhang, Anninos \& Norman}{1995}]{zan95}
Zhang Y., Anninos P., Norman M.~L., 1995, ApJ, 453, L57


\end{thebibliography}
\end{document}